# Conveyor Line Color Object Sorting using A Monochrome Camera, Colored Light and RGB Filters

Mason Petersen
masonp@mtu.edu

Brendon Lakenen
bjlakene@mtu.edu

Krishna Chavan
kchavan@mtu.edu

Pratik Waghmare
pkwaghma@mtu.edu

Aleksandr Sergeyev
avsergue@mtu.edu

Nathir Rawashdeh
narawash@mtu.edu

*Mechatronics, Applied Computing, Michigan Technological University*

*Abstract*—This paper tests the ability of a machine vision system with a monochrome camera to differentiate colored objects. The system is designed to autonomously and continuously sort colored objects of which the user specifies the desired color(s). The system uses camera color light filters and red-green-blue (RGB) color light emitting diode (LED) lights to aid the machine vision system in recognizing part contrast. Additionally the system is controlled by a Programmable Logic Controller (PLC) which is integrated with a Robot that is used to remove parts. The parts are fed into the workcell via a conveyor belt that is controlled by the PLC. The user has the ability to select the desired acceptable colors on both the HMI and the physical pushbuttons. An online video can be found here.

## I. Introduction

This paper represents a significant advancement in manufacturing and logistics processes. By incorporating multiple components such as the conveyor belt, robotic arm, vision system, sensors, and storage area, the system demonstrates a sophisticated level of automation as seen in Fig. 1. The integration of these components enables the efficient identification of objects based on their color and user-defined requirements. Central to the system's operation is the Programmable Logic Controller (PLC), which serves as the core controller. The PLC plays a pivotal role in orchestrating the actions of all components, ensuring seamless coordination and synchronization. Its ability to receive color information from the vision system and compare the user input enables dynamic decision-making in real-time. Moreover, the inclusion of a vision system enhances the system's adaptability and versatility.

By accurately identifying objects based on their color, the vision system facilitates precise sorting and selection according to user inputs. This capability streamlines the process, reducing manual intervention and improving overall efficiency. Additionally, the utilization of a robotic arm for object manipulation underscores the system's capabilities in handling diverse tasks autonomously. The robotic arm's agility and precision enable it to pick up and place objects with accuracy, contributing to the system's reliability and effectiveness. Furthermore, the incorporation of sensors to detect object presence enhances operational efficiency and safety. These sensors enable the system to monitor the flow of objects along the conveyor belt, ensuring smooth and uninterrupted operation. Overall, this paper represents a paradigm shift in automated sorting and reorganization processes. By leveraging advanced technologies and intelligent control mechanisms, this paper offers unparalleled efficiency, accuracy, and flexibility in handling complex tasks. As industries increasingly embrace automation to enhance productivity and streamline operations, systems like these exemplify the transformative potential of modern automation solutions. This project and similar ones have been developed through the Mechatronics education program at Michigan Technological University [1-4].

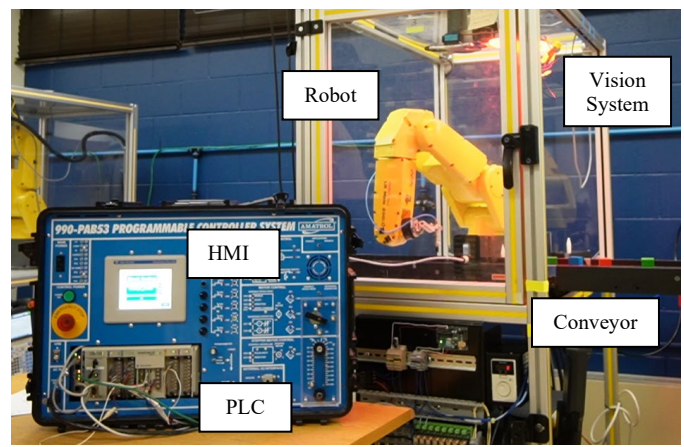

Fig. 1. Full system setup (PLC, HMI, robot, conveyor, and vision system)

## II. Literature Review

Previous works have tested machine vision systems' ability to differentiate color parts [5-8]. However, most use algorithms or non-monochrome cameras which allow for color detection. There does not seem to be any research on using color light filters in conjunction with a monochrome camera for color detection.

## III. System Architecture

The system process flow can be seen in Fig. 2 and operates as follows: The start button is pressed (physical or HMI) and the robot faults are cleared, which allows the conveyor to start and move forward. The conveyor is stopped when a part breaks the beam sensor line, this triggers the vision system to scan the part. The PLC will compare the user requested color to the detected color of the part and decide if the part is allowed to stay on the conveyor or if the part needs to be removed. If the part can stay then the conveyor turns on until the next part is detected, if the part needs to be removed then the robot locates the part, picks it up, and drops it off in the rejected part bin. Then the cycle continues until it is stopped by the operator. Additionally the system can be paused/held and then resumed just by using the stop and start buttons respectively.

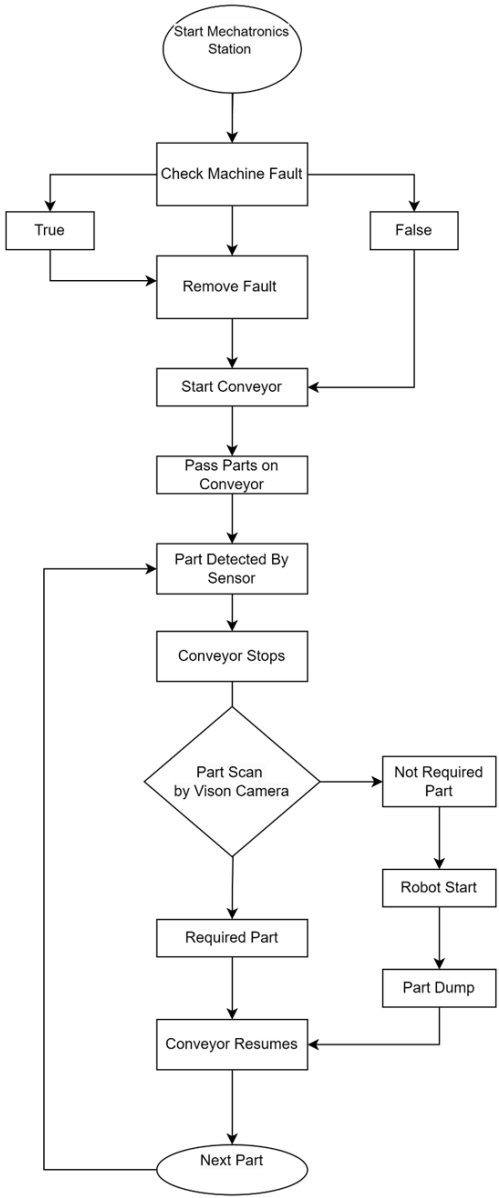

Fig. 2. System process flow diagram

### A. Monochrome Camera

The camera Fig. 3 is mounted at the top of the robot work cell, above the belt conveyor. It is a Sony XC - 56 monochrome camera and is used to supply images to the robotic vision system.

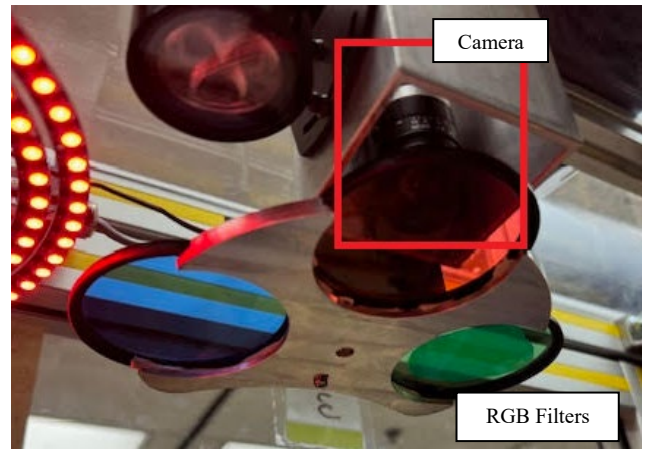

Fig. 3. Sony XC-56 monochrome camera (behind red lens)

### B. Color Light Filter Selector

The color light filter selector Fig. 4 is made up of red, green, and blue (RGB) optical filters in front of the camera lens. The filters are mounted to a custom made holding bracket which is connected to a 180 degree servo motor. The servo motor is mounted to a bracket that is attached to the top of the workcell in order to position the light filters in front of the camera.

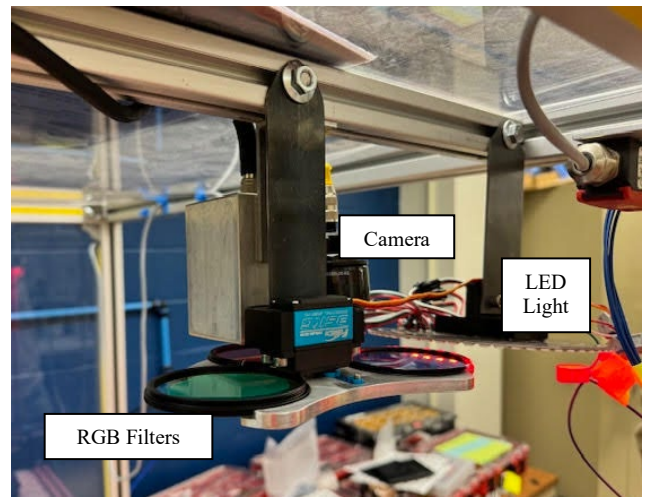

Fig. 4. Color light filter selector unit (servo, mounting bracket, filter holder, and filters)

### C. Color LED Disk

The Color RGB light emitting diode LED disk shown in Fig. 5 consists of 241 individually addressable LEDs that are oriented in a circle. The light can be configured to display any color in the RGB colorspace. This design creates a bright and even light, which helps the monochrome camera see contrast between colors.

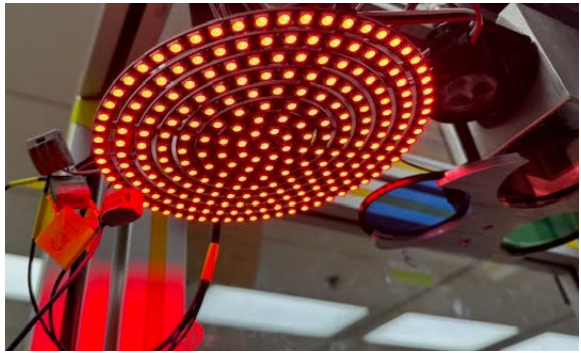

Fig. 5. Color LED disk light (set to red)

### D. Arduino

An Arduino Fig. 6 is an open source microcontroller that is equipped with Digital I/O pins. The Arduino is responsible for controlling the servo position and writing color values to the LEDs.

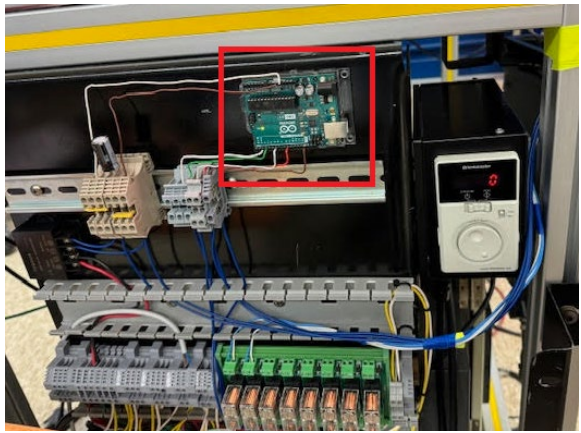

Fig. 6. Arduino unit that controls filter servo motor and LED lights

### E. Machine Vision System

The vision system used for this project is Fanuc's integrated IRVision Fig. 7. This system collects images from the monochrome camera and analyzes them.

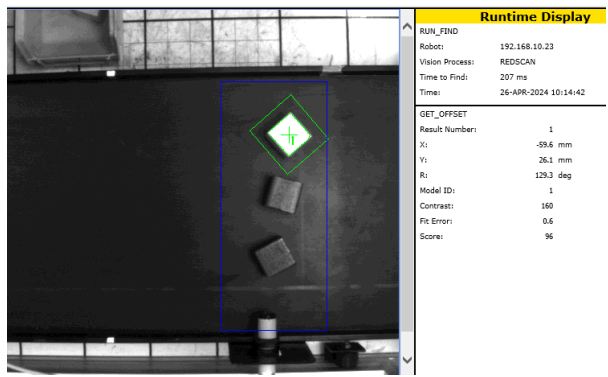

Fig. 7. Fanuc IRVision integrated development environment

It is capable of pattern and contrast detection to differentiate between parts. Additionally it can calculate the location of the part in the workcell, which can be used by the robot to pick the part. For example in the figure the parts from the top down are red, green, and blue respectively, here the vision system is only detecting the red one as a valid part. Which is correct as the vision inspection program is set up for the red part in this case. Additionally on the right side of the figure the X and Y axis translational data can be seen as well as the Z axis rotational data, this information is used by the robot to locate the part if it needs to be removed.

### F. Belt Conveyor

The belt conveyor Fig. 8 is used to transport products from the starting point to the camera search window. The belt conveyor has its own driver for speed and acceleration control but can be remotely turned on and off with a digital input.

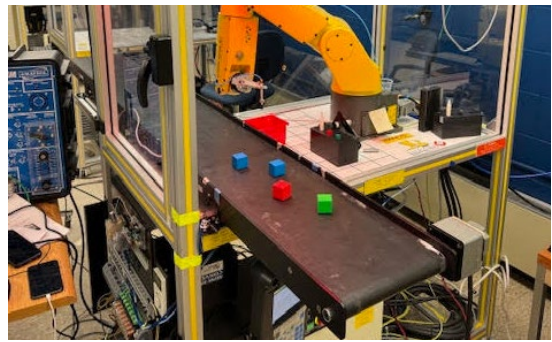

Fig. 8. Conveyor belt mounted in the robot workcell

### G. Robot

A Fanuc LR Mate 200ic Fig. 9 robot is used for removing parts that do not match the user requested color. The robot is equipped with a suction cup for EOAT that allows it to easily pick up blocks.

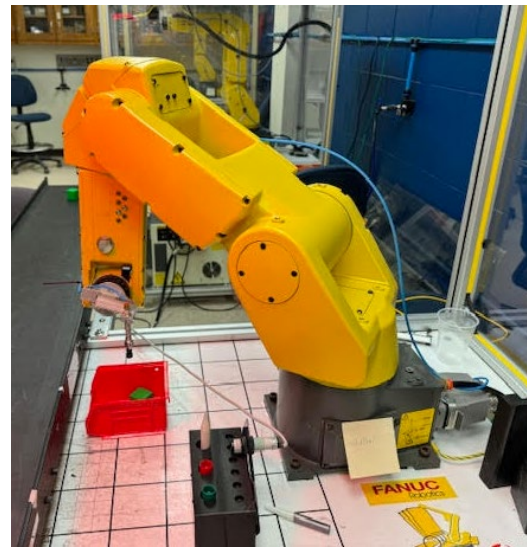

Fig. 9. Fanuc LR Mate robot equipped with a suction cup end effector.

### H. Programmable Logic Controller(PLC)

An Allen Bradley CompactLogix 5300 PLC Fig. 10 is used as the main system control. It is connected to the robot via ethernet and all other devices are wired to the I/O modules of the PLC.

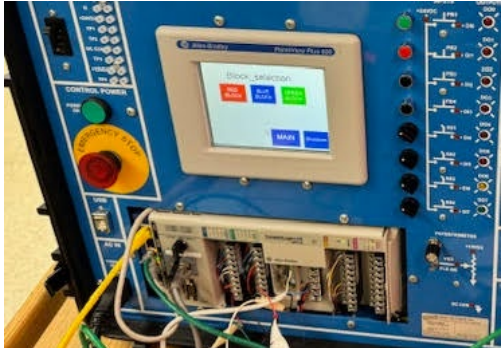

Fig. 10. Allen Bradley PLC and HMI

### I. Human Machine Interface(HMI)

The HMI used is an Allen Bradley PanelView Plus Fig. 10, this allows the user to select between block colors and control the start and stop of the system.

### J. Beam sensor

The beam sensors Fig. 11 are located on the conveyor belt, they are used to detect part presence, which the PLC uses to know when to stop the conveyor.

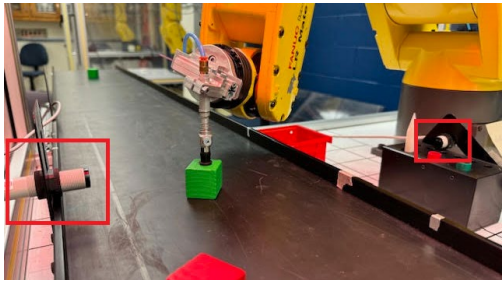

Fig. 11. Beam sensor mounted to conveyor to detect parts

### K. Reject Bin

The reject bin Fig. 12 is where the non-matching color parts are placed by the robot.

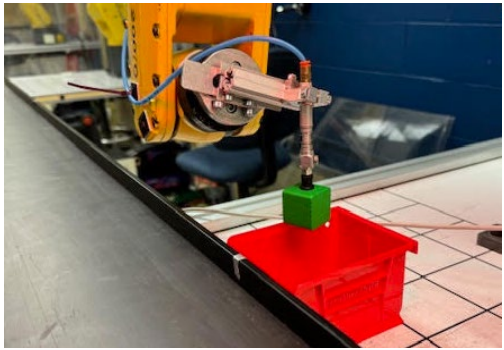

Fig. 12. Reject bin next to conveyor to store removed parts

## IV. METHODOLOGY

### A. Vision System Hardware

Monochrome cameras are not capable of differentiating color accurately as they only see brightness as seen in table 1. Thus using one to sort colors requires some modifications. Initially lens color filters were placed in front of the camera, in theory this would only allow light that matches the color of the filter to pass through to the camera. For example if a green, red, and blue block are placed on a dark background in front of the monochrome camera and a red filter is on the camera, then the resulting image should show the red block as light gray/white and the others as a dark gray/black. This would allow the vision system to pick out the red block by object contrast sorting, as it would have the greatest contrast compared to the background as seen in Fig. 14.

TABLE I. COLOR IMAGE VS MONOCHROME IMAGE

| | |
|---|---|
| Color Image | 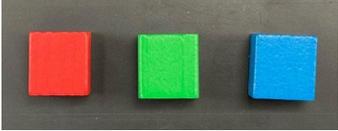 |
| Monochrome Image | 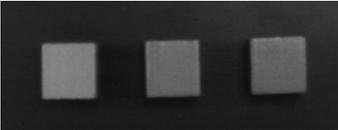 |

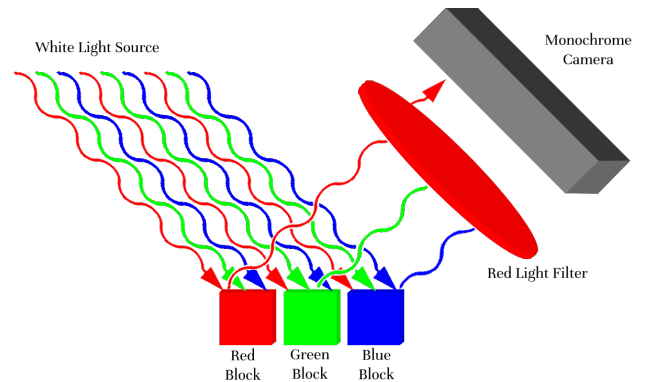

Fig. 13. Color light filtering to identify colors with a monochrome camera

In practice however the camera light filtering technique proved unreliable, red light filtering worked well enough to isolate the red block, but green and blue light filtering resulted in both the green and blue blocks appearing as light gray in the image, making them too hard to differentiate as seen in table 2. After further research it was determined that the color filters used were not of high enough quality to limit color wavebands to the necessary tolerance [9]. To resolve this supplementary color LED disk was used to flood the color parts with light and increase the contrast between colors in the images. In practice this created a significant enough contrast difference to discern part colors as seen in table 3.

TABLE II. MONOCHROME IMAGE WITH COLOR FILTERS.

| | |
|---|---|
| Red Filter | 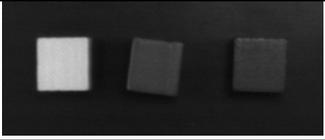 |
| Green Filter | 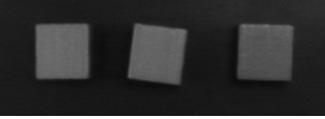 |
| Blue Filter | 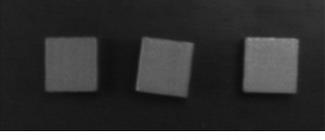 |
| Red, green, and blue blocks | |

TABLE III. MONOCHROME IMAGE WITH COLOR FILTER AND LED COLORED LIGHT

| | |
|---|---|
| Red Filter & Red LED | 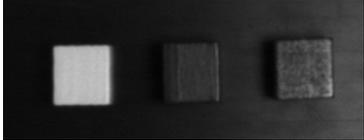 |
| Green Filter & Green LED | 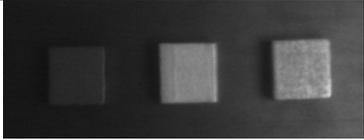 |
| Blue Filter & Blue LED | 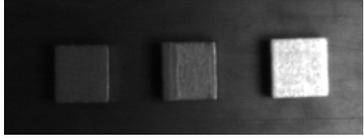 |
| Red, green, and blue blocks | |

Using the color light flooding and camera filtering the system is capable of identifying red, green, and blue parts. However, it can only identify one color at a time (per image). For example, to capture an image of a red part the system shines red light on the conveyor and the red filter is placed in front of the camera. When an image is taken the red part will appear white/light gray, the others will appear black/dark gray, and the conveyor will appear black. To capture images for blue and green parts the light and filter colors have to be changed to match the respective color. Therefore, scanning for red, green, and blue parts requires three separate images where the light and filter colors are changed for each. To sequence through the filters a servo motor with a filter holding fixture was implemented as seen in Fig. 4. The servo motor runs on 5 Volts and uses Pulse Width Modulation(PWM) for position control. The LED disk also runs on 5 Volts and uses a serial input to address and control the color values of the individual LEDs.

Because the robot and PLC use 24 Volt outputs, they are not suitable to directly control the servo and LEDs. Additionally, using a relay to jump from a 24 Volt output to a 5 Volt output is not an option for continuously changing signals like PMW and serial outputs as they operate at 50+ Hertz and will quickly wear out the relay. The solution to this was to use an Arduino which is capable of outputting 5 Volt PWM signals and serial signals. Furthermore, Arduino has libraries that are specifically designed for servomotor and addressable LED control [10-11].

*B. Vision System Program*

The program used on the Arduino for this project has two inputs and two outputs; the inputs are for color selection and the outputs are data signals to the servo and LEDs as seen in Appendix 1. If no input is true then the program defaults to output the red color, the servo will move to the angle that places the red filter in front of the camera and the LEDs will turn red As seen in Appendix 2. If the first input is true then the program outputs the green color and if the second output is true then the program outputs the blue color. The inputs come from the 24 Volt robot outputs, but they do so through a relay which isolates the 5 Volt circuit to protect the components. During initial testing a large amount of noise was noticed in the 5 Volt Arduino inputs which caused unpredictable system behavior. To resolve this the Arduino inputs needed to be defined as Pull Up resistor inputs which can be done in the Arduino program.

```
/PROG   SCANPART

   1:  DO[123:RED]=OFF
   2:  DO[124:GREEN]=OFF
   3:  DO[125:BLUE]=OFF
   4:
   5:  DO[110:CALL GREEN]=OFF
   6:  DO[112:CALL BLUE]=OFF
   7:  WAIT    .50(sec)
   8:  VISION RUN_FIND 'REDSCAN'
   9:  VISION GET_OFFSET 'REDSCAN' VR[1] JMP LBL[10]
  10:  DO[123:RED]=ON
  11:  LBL[10]
  12:
  13:  DO[110:CALL GREEN]=ON
  14:  WAIT    .50(sec)
  15:  VISION RUN_FIND 'GRNSCAN'
  16:  VISION GET_OFFSET 'GRNSCAN' VR[1] JMP LBL[20]
  17:  DO[124:GREEN]=ON
  18:  LBL[20]
  19:  DO[110:CALL GREEN]=OFF
  20:
  21:  DO[112:CALL BLUE]=ON
  22:  WAIT    .80(sec)
  23:  VISION RUN_FIND 'BLUSCAN'
  24:  VISION GET_OFFSET 'BLUSCAN' VR[1] JMP LBL[30]
  25:  DO[125:BLUE]=ON
  26:  LBL[30]
  27:  DO[112:CALL BLUE]=OFF
  28:
  29:  DO[130:SCAN COMPLETE]=ON
/END
```

Fig. 14. Fanuc robot scan part program

The Inputs to the Arduino are two of the robot's digital outputs which are controlled within one of the robot programs as seen in Fig. 14. The robot vision program functions as follows, first the robot turns off both digital outputs to the Arduino which moves the red filter in front of the camera and turns the LEDs red. Then the robot program calls its onboard vision system(FANUC IRVision) to scan the part. The vision system looks at the contrast difference between the part and the background. If there is a high contrast difference then the system identifies the part and saves the location data to a register on the robot controller. If the part is found the program will also turn on a digital output that tells the PLC it has identified a red part. This sequence is repeated for the green and blue color, in total three outputs are used from the robot to tell the PLC the color of the part and one final output is used at the end of the program to tell the PLC that the scanning is done.

*C. Allen Bradley PLC Ladder Logic*

Ladder logic is the most widely PLC language used when it comes to Robot programming. The PLC ladder logic was written in RSLogix 5000 Fig. 15. The ladder logic program controls the conveyor system, sends commands to the robot, and processes inputs from the sensor. It manages the conveyor's movement and instructs the PLC (programmable logic controller) to stop the conveyor when an object is detected by the proximity sensor. The PLC then signals the vision camera to inspect the detected object. If the object is acceptable, the PLC directs the conveyor to continue. However, if the object is unacceptable, the PLC commands the robot to remove the object from the conveyor and place it in a rejection bin.

To determine if an object is acceptable, the PLC compares the vision camera's input with the operator's predefined criteria, which could be a specific color block (red, green, or blue). If the camera input matches the operator's criteria, the PLC allows the conveyor to proceed and instructs the robot to stand by. Conversely, if the camera input does not match the criteria, the PLC orders the robot to remove the unacceptable object. The robot's actions are entirely dependent on the PLC's assessment of the camera input against the operator's standards.

Each rung in the ladder logic program serves a specific purpose. Rung 0 controls the conveyor's start and stop functions, utilizing HMI tags that enable operation through the human-machine interface panel. Activating the start button sets the Enable bit, which holds the robot's operation and signals the conveyor to begin running. Rungs 1 through 3 provide instructions for the robot, initiate vision camera scanning of the present block, and reset any robot faults. The <Local:1:I.Data.2> tag is assigned to the "Fault_Reset" in the PLC. In rung 5, the <Local:1:I.Data.14> and <ROBOT:I.Data[1].1> tags are normally closed but become enabled when rung 6 evaluates as true. Upon completing its scan, the vision camera sends a "scan done" output to the PLC, enabling <ROBOT:I.Data[1].1> and starting a timer delay. Rung 8 performs a comparison between the vision camera's output and the operator's input, evaluating as true if they match, <Local:1:I.Data.4>, <Local:1:I.Data.6>, <Local:1:I.Data.5>, <ROBOT:I.Data[0].10>, <ROBOT:I.Data[0].11>, and <ROBOT:I.Data[0].12> tags are normally open but become enabled when the PLC receives both the operator's instructions and the vision camera's scanned output, satisfying the respective rung conditions. This comparison operation evaluates the vision output against the PLC input, and if a match occurs, it enables the <ROBOT:O.Data[0].0> tag, indicating that the part confirms the PLC's input criteria.

If there's a discrepancy between the part and the inputs from both the PLC and the vision camera, it triggers the <ROBOT:O.Data[0].9> output tag, prompting the robot to execute its task. This process continues as long as the part remains detected and scanned by the vision camera.

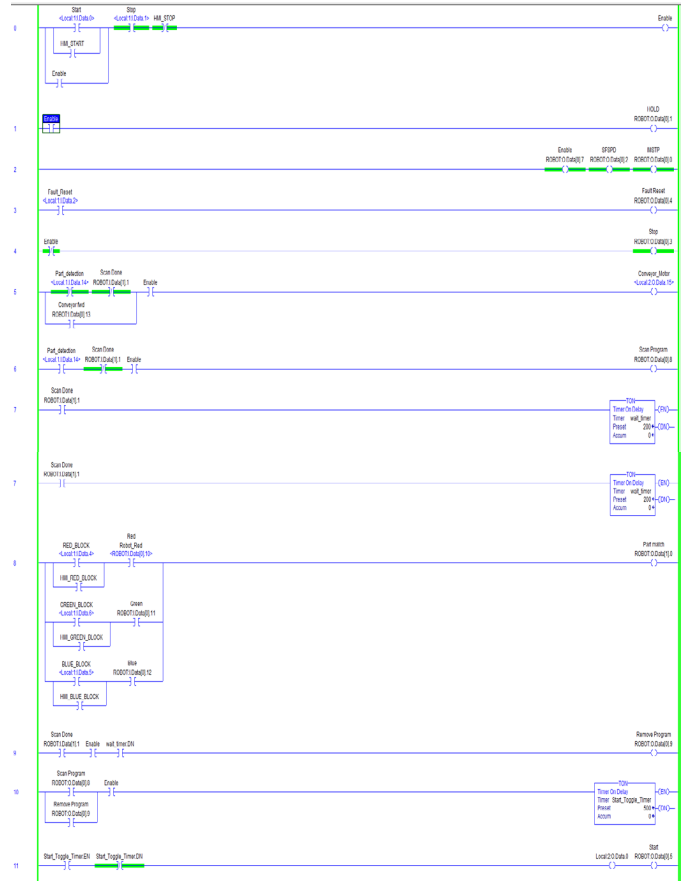

Fig. 15. PLC ladder logic

Figures 16 and 17 show the input and output tags for the robot to receive signals from the PLC and send signals back to the PLC for accurate operation. These input and output tags are important for the integration of the robot with the PLC. These tags change from 0 to 1 and vice versa as they turn ON and OFF. ROBOT:I.Data[0] and ROBOT:O.Data[0] tag is updated when the PLC provides input and if the corresponding part matches.

Fig. 16. The PLC's robot input tags

Fig. 17. The PLC's robot output tags

Figures 18 to 20 Show the HMI panel which is used to control the process. We have used HMI tags to integrate with PLC. By integrating PLC ladder logic we can use an HMI panel to run the conveyor and send the input.

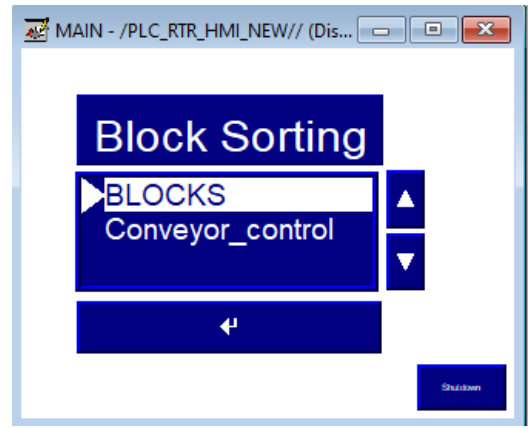

Fig. 18. HMI main screen

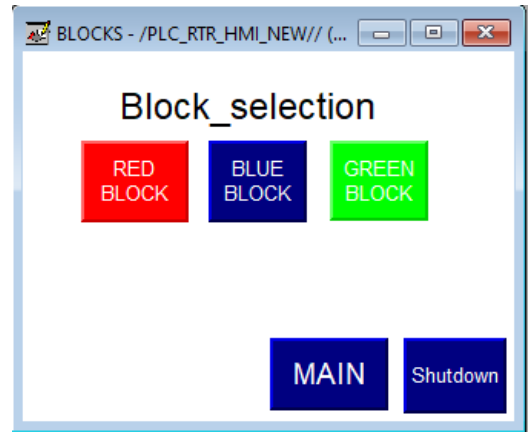

Fig. 19. HMI block selection screen

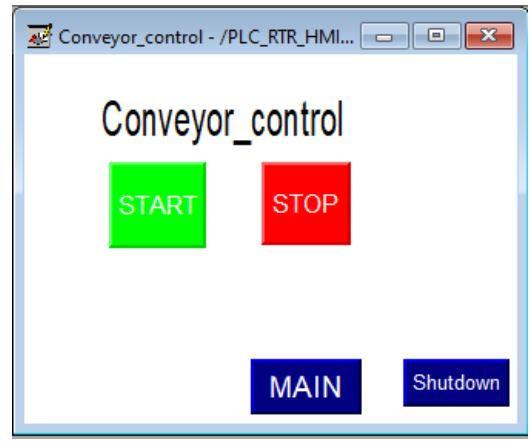

Fig. 20. HMI conveyor control screen

*D. Robot Sorting Program*

The robot sorting program Fig. 21 uses one digital input(from the PLC) to determine if a part is to be removed. If the input is on then the robot uses the part position data that is saved in the robot register to position the picking tool above the part. The tool moves down to pick the part up and then the robot carries the part to a reject bin next to the conveyor and drops the part. If the input is not on then the robot does not move, but the program toggles on a digital output for seconds, which controls the conveyor through the PLC and moves the part past the beam sensor. At the end of the program the color value outputs are all reset to off and the scan done output is set to off.

```
/PROG   SORTPART

   1:  UFRAME_NUM=8
   2:  UTOOL_NUM=8
   3:J P[1] 2% FINE
   4:
   5:   IF DI[121:REMOVE PART]=ON,JMP LBL[10]
   6:
   7:L PR[80:VISION REF] 100mm/sec FINE VOFFSET,VR[1]
    :  Offset,PR[81:Z_OFFSET]
   8:L PR[80:VISION REF] 50mm/sec FINE VOFFSET,VR[1]
   9:   WAIT    .50(sec)
  10:   DO[111:SUCTION CUP]=ON
  11:   WAIT    .50(sec)
  12:L PR[80:VISION REF] 50mm/sec FINE VOFFSET,VR[1]
    :  Offset,PR[81:Z_OFFSET]
  13:L P[1] 100mm/sec FINE Offset,PR[81:Z_OFFSET]
  14:L P[1] 75mm/sec FINE
  15:   WAIT    .50(sec)
  16:   DO[111:SUCTION CUP]=OFF
  17:   WAIT    .50(sec)
  18:
  19:   JMP LBL[11]
  20:   LBL[10]
  21:   DO[126:CONVEYOR FWD]=ON
  22:   WAIT    .75(sec)
  23:   DO[126:CONVEYOR FWD]=OFF
  24:
  25:   LBL[11]
  26:   DO[123:RED]=OFF
  27:   DO[124:GREEN]=OFF
  28:   DO[125:BLUE]=OFF
  29:   DO[130:SCAN COMPLETE]=OFF ;
/END
```

Fig. 21. Fanuc Robot Sort Part Program

## V. COMMUNICATION AND CONNECTION

### A. System Communication and Connection

The Allen Bradley PLC is connected to the Fanuc robot controller utilizing TCP/IP via ethernet cable. The Sony XC vision system is connected to the Fanuc robot through a PC, which is used to configure the vision system. Once the vision system is taught the program it will run directly off of the robot controller. The conveyor belt system is run off of 110 VDC supplied from the wall outlet and is controlled by the PLC via electrical wires with a 24VDC signal to exchange I/O's. There are three wires; input, output and ground. The two proximity sensors are connected to the PLC, the PLC provides power and control of the sensors. One of the proximity sensors is a four wire sensor that powers the sensor, provides input, and collects outputs from the other sensor and sends the data back to the PLC. The other sensor has two wires only used for power. Ethernet communication protocols, including a local area network (LAN) or a wide area network (WAN), are shared by all of the systems in order to have communication throughout the entire process. See Fig. 22 for the connection diagram.

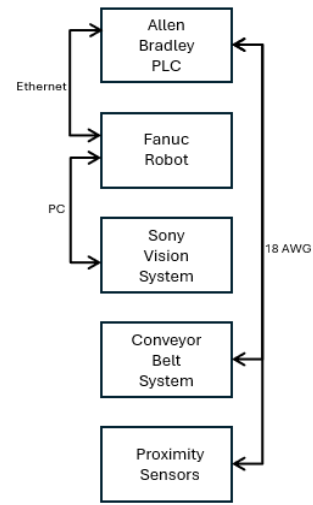

Fig. 22. System connection layout

### B. Communication and configuration between Robot and PLC

In order to enable communication between the Allen Bradley PLC and Fanuc robot utilizing ethernet connection the following procedure must be accomplished. The robot controller and the PLC must be updated to the required core software to ensure proper communication. The Robot will be programmed and connected to the PLC by the Teach Pendant. Under the setup window in the main menu you can set up Host communications utilizing TCP/IP Fig. 23. This is where you name the robot and assign IP addresses to the PC, PLC, Vision system and Robot. This is where the Subnet mask is assigned as well (255.255.255.0). After completing the host communications the PLC IP must be initiated and pinged to ensure proper communication. Once communication is provided the UI signals must be enabled and the Robot must be set to the REMOTE setup configuration. The next step is to set up and configure the PLC.

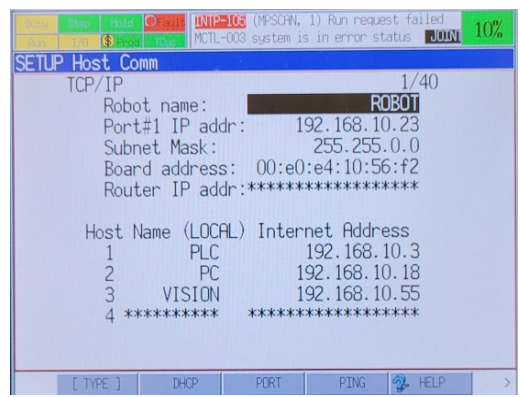

Fig. 23. Fanuc robot ethernet IP

### C. PLC Configuration Steps

The program used to program the PLC is RSLogix 5000. In order to complete communication between the PLC and robot the following procedure must be followed. Opening a new task in the main window is the first step required. Then a generic I/O Ethernet module must be generated. This is where the word length (4 for this project) , the input and output instances, and IP address of the robot Fig. 24 are assigned. Once the generic ethernet module is created check the controller tags to verify that the input/output tags were created. After verifying that everything is working properly write the ladder logic and configure I/O's for the program. See section V for Allen Bradley PLC programing instructions.

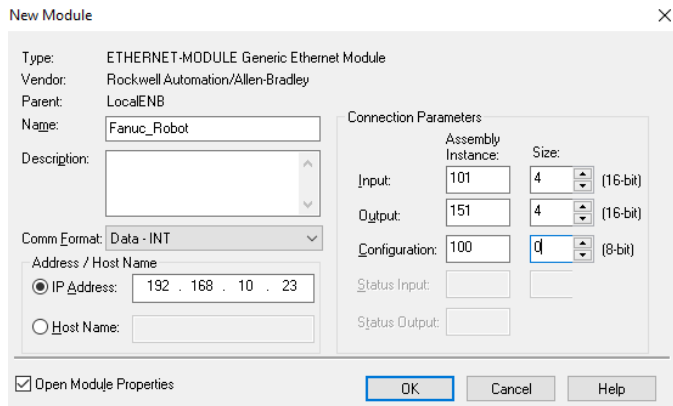

Fig. 24. PLC ethernet IP module

### D. Communication Between Robot and Vision system

The Sony vision system is connected to the Fanuc Robot via PC. Once the vision system is taught the required program it will be run off of the Fanuc Robot controller. The robot in the lab has a built in vision system which allows an easy communication process. If there was not a built in vision system you could integrate one via the robot controller and program it utilizing a PC. The vision system sends feedback to the PLC in order to execute the proper tasks required during operation.

## VI. RESULTS

The system proved to be capable of continuous color object sorting with minimal errors. Fanuc's IRVsion system accurately detects the trained colored blocks and the PLC correctly sorts the colors of the blocks according to the user specified input. Integrating both color filtration and supplementary light is crucial to yield a good contrast between the color object and the background.

## VII. DISCUSSION

It should be noted that the system did have some minor limitations that would need to be addressed in-order for the system to be integrated into an industrial application. As seen on table 3 the filtering and supplemental lighting method works very well for depicting red and blue objects. However, the results for the green filter and light are not as distinct. The green part has a slightly higher contrast than the blue block does, which is enough for the machine vision system to differentiate them when they are in the middle of the conveyor belt. But when the blocks are closer to the edge of the conveyor belt(non-centered to the camera) the vision system will confuse the blue block for the green block. This study did not have time to investigate further lighting and filtering options to resolve the issue, but proposed that an additional statement could be made to the program to resolve the issue. This being that the scan program will already accurately identify blue parts with a blue light and blue filter, so if a blue part is falsely identified as green then both the green and blue outputs to the PLC will be on. A program statement that could fix this would look as follows:

If Green & Blue outputs = ON
⇨ The Blue Part is present, but has been falsely identified as green, so leave Blue output on and turn off Green output.

Another system error noticed was with the vision systems part locating ability. Occasionally the suction cup would not be centered on the top of the block to pick up the part like it is meant to be. There was not enough time to research this in this study, but possible culprits could be: movement in the camera location due to moving mass of filter holding fixture, light refraction through the camera filters, or inaccuracies in the vision training sequence.

ACKNOWLEDGMENT

This project is made possible with the help of Prof. Rawashdeh, Prof. Sergeyev, and the facilities at Michigan Technological University - Department of Applied Computing.

APPENDIX

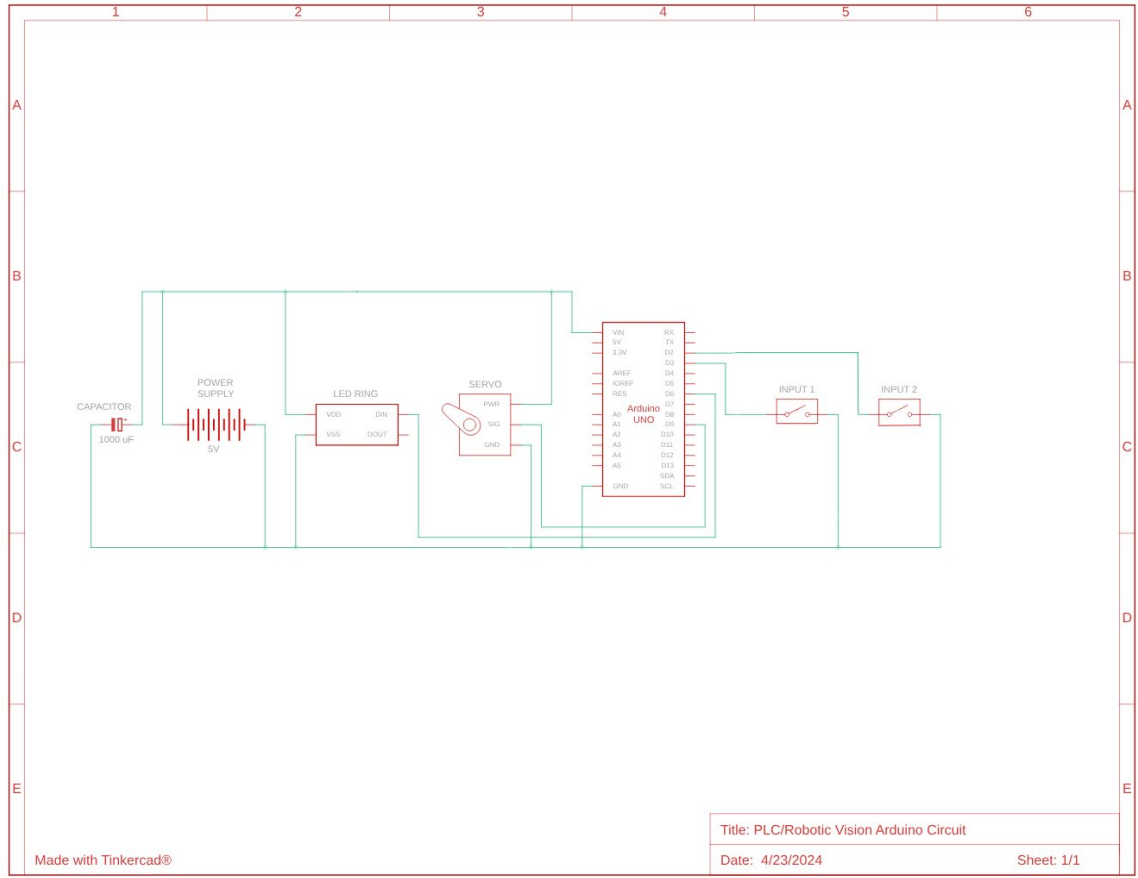

Appendix. 1. Arduino circuit wiring

```cpp
#include <Adafruit_NeoPixel.h>
#include <Adafruit_TiCoServo.h>

#define N_LEDS      241
#define LED_PIN       6

#define SERVO_PIN    9
#define SERVO_MIN    500
#define SERVO_MAX    2500

#define Input_1      2
#define Input_2      3

int angle;
int red;
int green;
int blue;

Adafruit_NeoPixel  strip = Adafruit_NeoPixel(N_LEDS, LED_PIN);
Adafruit_TiCoServo servo;

void setup(void) {
  pinMode(Input_1, INPUT_PULLUP);
  pinMode(Input_2, INPUT_PULLUP);
  servo.attach(SERVO_PIN, SERVO_MIN, SERVO_MAX);
  strip.begin();
}
void loop(void) {

int  A = digitalRead(Input_1);
int  B = digitalRead(Input_2);

  if (A == LOW)
  {
    angle = 180;
    red = 0;
    green = 255;
    blue = 0;
  }

  else if (B == LOW)
  {
    angle = 0;
    red = 0;
    green = 0;
    blue = 255;
  }
  else
  {
    angle = 90;
    red = 255;
    green = 0;
    blue = 0;
  }

  servo.write(angle);

  for(int i=0; i<N_LEDS; i++) { // For each pixel...

    strip.setPixelColor(i, strip.Color(red, green, blue));
  }
  strip.show();
}
```

Appendix. 2.      Arduino program

# BIOGRAPHIES

BRENDON LAKENEN is a Graduate student in the department of Applied computing at Michigan Tech. since August 2023. He is working on his Masters in Mechatronics and certification in Fluid Power. Prior to this, he was an undergraduate student in the Mechanical Engineering Technology bachelor's program at Michigan Tech. , where he spent 4 years. His experience during his time at Michigan Tech. includes 4 years with the Aerospace Enterprise.

PRATIK WAGHMARE is a Graduate student in the department of Applied computing at Michigan Tech. since August 2023. He is a graduate student pursuing his degree in Mechatronics. He has a Bachelor of Technology degree in mechanical engineering, from Maharashtra Institute of Technology, Aurangabad. After graduation he worked as a trainee engineer in AITG industries and also did a summer internship at SPAL Automotive.

MASON PETERSEN is a Graduate student in the department of applied computing at Michigan Technological University and expected to graduate in April of 2024 with a Masters degree in Mechatronics. Mason received his undergraduate degree in Manufacturing & Mechanical Engineering Technology also at Michigan Technological University in April of 2023. After graduation Mason is going to work as a controls engineer at Axis Automation in Walker, Michigan.

KRISHNA CHAVAN is a graduate student in the Department of Applied Computing at Michigan Technological University, since August 2023. He is pursuing a degree in Mechatronics. He holds a Bachelor of Technology degree in Mechanical Engineering from Pune University, India. After graduation, he worked as a trainee engineer at Ognibene India Pvt Ltd for two years and as an Information Systems Engineer at Infosys Limited for one year.

ALEKSANDR SERGEYEV
Aleksandr Sergeyev is a professor in the College of Computing at Michigan Tech. He is also an affiliate professor in the Electrical and Computer Engineering department. His areas of expertise are electrical and computer engineering, physics, and adaptive optics. Sergeyev's professional interests include robotics, and he is a certified instructor for the FANUC Robotic Automation Industrial Certification through the Certified Education Robot Training (CERT) Program. He is also involved with the Michigan Tech Child Development Advisory Board and the Student Commission on campus.

NATHIR RAWASHDEH is an assistant professor at the department of applied computing at Michigan Tech. since August 2019. He instructs introductory and advanced programmable logic controller courses. Prior to this appointment, he was an associate professor in the Mechatronics Engineering Department at the German Jordanian University, where he spent 10 years. His industrial experience includes 5 years with Lexmark International, Inc. Lexington-Kentucky and MathWorks, Inc. in Natick-Massachusetts. Dr. Rawashdeh is a Senior Member of the IEEE and has experience with European-funded research capacity building projects. His research interests include mobile robots, autonomous driving, image processing and sensor fusion.